\PassOptionsToPackage{table,xcdraw}{xcolor}
\documentclass[11pt,a4paper]{article}
\pdfoutput=1

\makeatletter
\def\@fpheader{}
\makeatother

\usepackage{jheppub}
\usepackage{gensymb}
\DeclareSymbolFont{matha}{OML}{txmi}{m}{it}
\DeclareMathSymbol{\varv}{\mathord}{matha}{118}
\usepackage{amsmath}
\usepackage{amssymb}
\usepackage{graphicx}
\usepackage{subfigure}
\usepackage{pstricks}
\usepackage{bm}
\usepackage{pbox}
\usepackage{placeins}
\usepackage{booktabs}
\usepackage[T1]{fontenc}
\usepackage{footnote}
\usepackage{pdfpages}
\usepackage{hhline}
\usepackage{multirow}
\usepackage{multicol}
\usepackage{enumitem}
\usepackage[toc,page]{appendix}
\allowdisplaybreaks
\usepackage{comment}
\usepackage{float}
\usepackage{slashed}
\usepackage{xcolor}
\usepackage{textcomp}
\usepackage{multirow}
\usepackage[numbers]{natbib}
\usepackage{notoccite}
\usepackage{indentfirst}
\usepackage{jheppub}
\usepackage{gensymb}
\DeclareSymbolFont{matha}{OML}{txmi}{m}{it}
\DeclareMathSymbol{\varv}{\mathord}{matha}{118}
\usepackage{amsmath}
\usepackage{amssymb}
\usepackage{graphicx}
\usepackage{subfigure}
\usepackage{pstricks}
\usepackage{bm}
\usepackage{pbox}
\usepackage{placeins}
\usepackage{booktabs}
\usepackage[T1]{fontenc}
\usepackage{footnote}
\usepackage{pdfpages}
\usepackage{hhline}
\usepackage{multirow}
\usepackage{multicol}
\usepackage{enumitem}
\usepackage[toc,page]{appendix}
\allowdisplaybreaks
\usepackage{comment}
\usepackage{float}
\usepackage{slashed}
\usepackage{xcolor}
\usepackage{textcomp}
\usepackage{multirow}
\usepackage[numbers]{natbib}
\usepackage{notoccite}
\usepackage{soul}
\usepackage{extarrows}

\usepackage{physics}
\usepackage{comment}
\usepackage{dsfont}
\usepackage{blindtext}
\usepackage{csquotes}
\usepackage{soul}
\usepackage[none]{hyphenat}
\usepackage{indentfirst}
\DeclareUnicodeCharacter{0304}{ }
\usepackage{rotating}
\usepackage{tabularx}
\usepackage{extarrows}
\usepackage{physics}
\usepackage{comment}
\usepackage{dsfont}
\usepackage{blindtext}
\usepackage{csquotes}
\usepackage[none]{hyphenat}
\DeclareUnicodeCharacter{0304}{ }
\usepackage{rotating}
\usepackage{tabularx}

\usepackage[many]{tcolorbox}
\definecolor{fg}{RGB}{34,139,34}

\renewcommand{\thetable}{\Roman{table}}

\def\figureautorefname~#1\null{Fig.\,#1\null}
\def\equationautorefname~#1\null{Eq.\,(#1)\null}
\def\tableautorefname~#1\null{Tab.\,#1\null}
\definecolor{MyDarkBlue}{rgb}{0.1, 0.1, 0.8} 
\definecolor{MyLightBlue}{rgb}{0.22,0.51,0.9}
\definecolor{MyGreen}{rgb}{0.0, 0.5, 0.0}
\definecolor{BrickRed}{rgb}{0.8, 0.25, 0.33}

\RequirePackage{hyperref}
\hypersetup{colorlinks=true, citecolor=MyGreen,linkcolor=BrickRed, urlcolor=MyLightBlue}

\title{\bf Gravitational Wave Signatures of a Chiral Fermion\\ Dark Matter Model}
\author[a]{Tomohiro Abe,}
\author[b]{K.S. Babu,}
\author[b]{Ajay Kaladharan}
\affiliation[a]{Department of Physics, Faculty of Science and Technology, Tokyo University of Science, Noda,
Chiba 278-8510, Japan}
\affiliation[b]{Department of Physics, Oklahoma State University, Stillwater, OK 74078, USA}
\emailAdd{abe.tomohiro@rs.tus.ac.jp}\emailAdd{babu@okstate.edu}\emailAdd{kaladharan.ajay@okstate.edu}

\abstract{Theories in which the dark matter (DM) candidate is a fermion transforming chirally under a gauge symmetry are attractive, as the gauge symmetry would protect the DM mass. In such theories, the universe would have undergone a phase transition at early times that generated the DM mass upon spontaneous  breaking of the gauge symmetry. In this paper, we explore the gravitational wave signals of a simple such theory based on an $\mathrm{SU}(2)_\mathrm{D}$ dark sector with a dark isospin-$3/2$ fermion serving as the DM candidate. This is arguably the simplest chiral theory possible.  The scalar sector consists of a dark isospin-$3$ multiple, which breaks the $\mathrm{SU}(2)_\mathrm{D}$ gauge symmetry and also generates the DM mass. We construct the full thermal potential of the model and identify regions of parameter space which lead to detectable gravitational wave signals, arising from a strong first-order $\mathrm{SU}(2)_\mathrm{D}$ phase transition, in various planned space-based interferometers,  while also being consistent with dark matter relic abundance. The bulk of the parameter space exhibiting detectable gravitational wave signals in the model also has large WIMP-nucleon scattering cross sections, $\sigma_{\rm SI}$, which could be probed in upcoming direct detection experiments.
}

\begin{document}
\maketitle

\begin{sloppypar}

\section{Introduction}
\label{sec:intro}
The first direct gravitational wave (GW) detection was achieved in 2015 by the LIGO  and Virgo collaborations~\cite{LIGOScientific:2016aoc}, which detected faint transient signals in the frequency range of $(35-250)$ Hz, attributed to binary black hole mergers. Within two years of the initial discovery, these collaborations also discovered gravitational waves from binary neutron star mergers with a gamma-ray-burst counterpart~\cite{PhysRevLett.119.161101}. Proposed next-generation detectors~\cite{Caprini:2015zlo,Caprini:2019egz, Corbin:2005ny,Kudoh:2005as}  could probe GW signals in a broader frequency spectrum in the range  $(10^{-4}-10^4)\,\mathrm{Hz}$, which would be also sensitive to primordial gravitational waves generated in the early history of the universe.  Discovery of such stochastic GWs, resulting from cosmological first-order phase transitions (FOPT), could serve as an important tool to probe the imprint of new physics in the early universe.  Even though these signals are too weak to be observed by ground-based gravitational-wave detectors~\cite{LIGOScientific:2016aoc}, they could be probed in planned space-based interferometers such as LISA~\cite{Caprini:2015zlo,Caprini:2019egz}, BBO~\cite{Corbin:2005ny} and DECIGO~\cite{Kudoh:2005as}. 
Primordial GW signals that have survived to the present day should have been produced by a mechanism that requires physics beyond the standard model (BSM).
The electroweak phase transition within the Standard Model (SM) is not strongly first order~\cite{Kajantie:1995kf,Kajantie:1996mn}, nor is the QCD phase transition~\cite{Aoki:2006we,Bhattacharya:2014ara}; so the detection of primordial GW signals would be an indication of BSM physics. In this backdrop, GWs emerging from electroweak symmetry breaking in extended models~\cite{Baldes:2018nel,  Alves:2018oct,  Fujikura:2018duw, Beniwal:2018hyi,  Addazi:2018nzm, Alves:2018jsw, Athron:2019teq, Bian:2019kmg, Wang:2019pet, Alves:2019igs, Goncalves:2021egx, Goncalves:2023svb, Goncalves:2022wbp, Ellis:2018mja, Alanne:2019bsm, Shajiee:2018jdq, Paul:2019pgt, Cao:2022ocg, Shibuya:2022xkj, Chakrabarty:2022yzp, Benincasa:2022elt, Arcadi:2022lpp} and in high scale breaking of new  symmetries~\cite{Okada:2018xdh, Prokopec:2018tnq, Brdar:2018num,  Marzo:2018nov,  Hasegawa:2019amx, Addazi:2019dqt, Huang:2020bbe, Graf:2021xku} have been analyzed by various authors. The goal of this paper is to analyze such signals in a model with chiral fermion dark matter (DM), which is motivated on the ground that the DM mass cannot take arbitrarily large values, as it arises from the Higgs mechanism.

Weakly interacting massive particles (WIMPs)  have been very popular DM candidates. Typically, in these models, DM is thermally produced in the early universe, and as the universe cools, at some temperature $T \sim m_{\rm DM}/20$, its interaction rates with the plasma become weaker than the Hubble expansion rate, whence the DM goes out of thermal equilibrium and freezes out.  This mechanism can explain the relic abundance of DM for a wide range of DM mass, $m_{\rm DM} \sim$ sub-GeV to a few tens of TeV, with appropriate cross sections for them to annihilate into SM particles.
The WIMP dark matter models typically treat the mass of the DM to be an input which is then fitted to obtain the correct relic abundance.  This would, however,  leave the scale of the DM mass unexplained.  In particular, in most instances, the DM mass is allowed theoretically to be as large as the Planck scale.

There is an interesting class of WIMP dark matter models where the DM particle is a fermion transforming chirally under a gauge symmetry. Its mass should then arise from spontaneous gauge symmetry breaking. Such ``chiral fermion DM models'' are attractive as they would also explain why the DM mass is not arbitrarily large. An example of such a model would be a fourth-generation Dirac neutrino dark matter with its mass protected by the Standard Model gauge symmetry.  However, this model has been ruled out by a combination of direct detection limits~\cite{XENON:2018voc} and 
collider search limits. If the chiral gauge symmetry operates in the dark sector, there would be no conflict with direct detection constraints, or collider constraints. A simple such theory was proposed in Ref.~\cite{Abe:2019zhx} by two of the present authors based on an $\mathrm{SU}(2)_\mathrm{D}$ dark sector gauge symmetry.  The DM candidate belongs to a dark isospin-3/2 fermion ($\psi$) of this model.  Such a theory is chiral in the sense that a mass term for $\psi$ cannot be written down, and its mass should arise from Yukawa interactions with a scalar field that spontaneously breaks the gauge symmetry.  
Other examples of chiral fermion dark matter models include mirror models \cite{Berezhiani:1995yi,Berezhiani:2000gw,Foot:1991bp,Foot:2004pa}, models which use incomplete mirror copies of the Standard Model fermions \cite{Berryman:2016rot,Berryman:2017twh} and variants of the left-right symmetric model \cite{BhupalDev:2016gna,Dror:2020jzy}. 

It should be noted that chiral gauge theories are strongly constrained since they should be free of chiral anomalies. While $\mathrm{SU}(2)_\mathrm{D}$ theories with any fermion representation has no chiral anomaly, it should be also free from global Witten anomaly \cite{Witten:1982fp}. This is determined by the quadratic index of the fermion representation. 
The quadratic index $\mu$ of representation $I$ in $SU(2)$ is given by $\mu = 2I(I + 1)(2I + 1)/3$. If $\mu$ is an odd integer, there is a global anomaly, as happens with a single $I=1/2$ fermion for which $\mu = 1$, while for $\mu$ even, there is no global anomaly. The simplest choice leading to a chiral theory with no global anomaly is $I =3/2$, which gives $\mu=10$.  (Note that the case of $I=1$ leads to a vector-like theory since a bare ass term is admissible for the fermion in this case.) This simple chiral model was first studied in the context of dynamical supersymmetry breaking in Ref.~\cite{Intriligator:1994rx}, and this model has been applied to generate naturally light sterile neutrinos in Ref.~\cite{Babu:2003is}.  In Ref.~\cite{Abe:2019zhx} such a chiral model was constructed with a dark isospin-3 scalar field utilized for spontaneous symmetry breaking and DM mass generation. 
The models developed in Ref.~\cite{Intriligator:1994rx,Babu:2003is,Abe:2019zhx} are perhaps the simplest chiral gauge theory.  If one were to construct a chiral gauge theory based on a $U(1)$ symmetry, at least five Weyl fermion fields would be needed for anomaly cancellation, viz;, $\sum_i Q_i = 0$ and $\sum_i Q_i^3 = 0$, from the absence of mixed gravitational anomaly and the $U(1)^3$ anomaly.  Examples include five Weyl fermions having integer charges given by $(23,-14,54,-55,-8)$ \cite{Babu:2003is}. For alternative choices of chiral $U(1)$ charges and for other solutions to the anomaly conditions see Ref.~\cite{Sayre:2005yh,Batra:2005rh,Costa:2019zzy,Costa:2020dph}.

The spontaneous breaking of a dark sector gauge symmetry in the early universe could very well be from a first-order phase transition.  In this case, the dark sector could be accessible through gravitational wave signatures. GW signals arising from dark sector phase transition with gauge symmetry assumed to be $U(1)$~\cite{Madge:2018gfl, Kannike:2019wsn, Mohamadnejad:2019vzg, Kannike:2019mzk, Alanne:2020jwx, Ertas:2021xeh, Wang:2022akn, Costa:2022oaa, Arcadi:2023lwc, Kanemura:2023jiw, Hosseini:2023qwu,Bian:2019szo}, $SU(2)$~\cite{Abe:2023zja, Baldes:2018emh, Fairbairn:2019xog, Ghosh:2020ipy}, and $SU(3)$~\cite{Helmboldt:2019pan, Dunsky:2019upk} have been examined and model agnostic analyses were carried out in Ref.~\cite{Dent:2022bcd, Croon:2018erz}. The stability of DM is guaranteed in these theories by discrete remnants of the gauge symmetry. However, GW signals arising from chiral fermion dark matter models have not been studied thus far, which is the focus of this work.  We investigate the phase transition pattern and the associated GWs in the chiral fermion DM model of Ref.~\cite{Abe:2019zhx}, which also provides a stabilization mechanism for the DM.  We have constructed the full one-loop finite temperature effective potential for the model, and carried out a systematic numerical scan of the parameters of the model to search for observable gravitational wave signals, while also being consistent with DM relic abundance. We find ample points in the parameter space that satisfy these criteria.  In regions where GW signals are observable, the DM-nucleus cross-section tends to be relatively large, which suggests that the model may be explored in direct detection experiments as well.

The rest of the paper is organized as follows. In~\autoref{sec:model}, we discuss the essential features of the model relevant for the study of phase transitions. 
In~\autoref{sec:constraints}, we summarize the various theoretical and experimental constraints on the model. 
In~\autoref{sec:effpot}, we derive the one-loop finite-temperature effective potential. A discussion on phase transition follows it in~\autoref{sec:phase_transition}. In~\autoref{sec:GW}, we focus on the gravitational wave signals associated with the phase transition in the dark sector. In Sec. \ref{sec:numer} we present our numerical analysis for the GW signals and confront it with the sensitivities of the planned experiments, and we conclude in~\autoref{sec:conclusion}. 
\section{The Model}
\label{sec:model}
Here we briefly review the chiral fermion dark matter model of Ref.~\cite{Abe:2019zhx}.  It is based on the dark sector gauge symmetry $\mathrm{SU}(2)_\mathrm{D}$ with a dark isospin-3/2 fermion $\psi$ denoted as
\begin{equation}
\psi^t
=
\begin{pmatrix}
 \psi_{3/2},\ \psi_{1/2},\ \psi_{-1/2},\ \psi_{-3/2}
\label{eq:4-plet-for-SU2}
\end{pmatrix}.
\end{equation}
As discussed in the introduction, this choice of fermion representation leads to a consistent theory, as it has no global $SU(2)$ anomaly. 
The group multiplication rule
\begin{equation}
\frac{3}{2} \otimes \frac{3}{2} = 0_{ a} \oplus 1_{ s} \oplus 2_{\rm a} \oplus 3_{ s}~,
\end{equation}
where the subscript $a$ ($s$) stands for antisymmetric (symmetric) combination, shows that a bare mass term for $\psi$ is not gauge invariant with a single $\psi$ field.  Furthermore, it suggests that dark isospin-1 or isospin-3 scalar should be introduced to generate a mass for $\psi$.  Since the isospin-1 choice would leave an unbroken $U(1)$, thereby leading to dark radiation, which is highly constrained from CMB measurements, the model introduces a dark isospin-3 scalar field, a $7$-plet $\phi$, denoted as:
\begin{equation}
\phi^t
=
 \begin{pmatrix}
  \phi_3,\   \phi_2,\   \phi_1,\   \phi_0,\ \phi_{-1},\   \phi_{-2},\   \phi_{-3} 
 \end{pmatrix}.
\end{equation}
Using the self-duality property of the $7$-plet, we identify the components of $\phi$ as:
\begin{align}
 \phi_{-3} = - \phi_3^*,\quad
 \phi_{-2} = + \phi_2^*,\quad
 \phi_{-1} = - \phi_1^*,\quad
 \phi_{0} =  + \phi_0^*.
\end{align}

\subsection{Yukawa couplings}
\label{sec:Yukawa}
The Yukawa couplings of the fermion $\psi$ with the scalar $\phi$ can be written explicitly as:
\begin{align}
{\cal L}_{\mathrm{Yukawa}
}=- y_D&\left [ \phi_{-3} \psi_{3/2} \psi_{3/2}
- \sqrt{2} \phi_{-2} \psi_{1/2} \psi_{3/2}
+ \phi_{-1}\left( \sqrt{\frac{3}{5}} \psi_{1/2} \psi_{1/2} + \frac{2}{\sqrt{5}} \psi_{3/2} \psi_{-1/2}  \right)  \right. \nonumber\\
&- \phi_0 \left( \frac{3}{\sqrt{5}} \psi_{1/2} \psi_{-1/2} + \frac{1}{\sqrt{5}} \psi_{3/2} \psi_{-3/2} \right)- \sqrt{2} \phi_{2} \psi_{-1/2} \psi_{-3/2}
+\phi_{3} \psi_{-3/2} \psi_{-3/2}  \nonumber\\
&\left. + \phi_1 \left( \sqrt{\frac{3}{5}} \psi_{-1/2} \psi_{-1/2} + \frac{2}{\sqrt{5}} \psi_{-3/2} \psi_{1/2}  \right) \right ]+\mathrm{h.c}. 
\label{Yuk}
\end{align}
As we shall see, spontaneous symmetry breaking occurs in the model when the component $\phi_2$ acquires a nonzero vacuum expectation value (VEV), in which case all four Weyl components of $\psi$ would acquire a common mass from Eq. (\ref{Yuk}).

\subsection{Higgs potential analysis}
\label{sec:potential}
The scalar potential of the model is given by~\cite{Abe:2019zhx}
\begin{align}
 V_{\mathrm{scalar}} =&\mu_H^2 H^\dagger H 
+ \lambda (H^\dagger H)^2 
+ m^2 \left( \frac{1}{2} \phi_0^2 + |\phi_1|^2 + |\phi_2|^2  + |\phi_3|^2 \right)
\nonumber\\
&+ \lambda_{H\phi} H^\dagger H \left( \phi_0^2 + 2 |\phi_1|^2 + 2 |\phi_2|^2  + 2 |\phi_3|^2 \right)
+ \lambda_1 \left( \frac{1}{2} \phi_0^2 + |\phi_1|^2 + |\phi_2|^2  + |\phi_3|^2 \right)^2 \nonumber
\\
& + \lambda_2 
\Biggl(
\left| \sqrt{6} \phi_1^2 - 2 \sqrt{5} \phi_0 \phi_2 + \sqrt{10} \phi_3 \phi_{-1} \right|^2
+ 
\left| \sqrt{2} \phi_0 \phi_1 - \sqrt{15} \phi_2 \phi_{-1} + 5 \phi_3 \phi_{-2} \right|^2 \nonumber
\\
&+ \frac{1}{2}
\left( 2 \phi_0^2 + 3 |\phi_1|^2  - 5 |\phi_3|^2 \right)^2
\Biggr)
,
\end{align}
%
%
%
where $H$ is the SM Higgs doublet. We choose the mass parameters $(m^2,\,\mu_H^2)$ to be both negative, so that at zero temperature the ground state of the theory would be away from 
$\langle \phi \rangle = 0$ and $\langle H \rangle = 0$.  At $T=0$ we also assume, following Ref.~\cite{Abe:2019zhx}, that Im($\phi_2$) acquires a nonzero VEV.  Expanding around the VEVs, the $H$ and $\mathrm{Im}\phi_2$ fields can be written as
\begin{align}
    H = \left(\begin{array}{c}
        G^+\\
        \frac{1}{\sqrt{2}}\left(h + \sigma + i G^0\right)
    \end{array}\right),
    \hspace{0.5cm} 
    \hspace{0.5cm}
        \mathrm{Im}\phi_2 =- \frac {s+\sigma_7}{\sqrt{2}},
        \label{eq:expand}
\end{align}
where at zero temperature we define $h|_{T=0} \equiv v$ and $s|_{T=0} \equiv v_D$, with $v = 246$ GeV.

A $7$-plet scalar of $\mathrm{SU}(2)_\mathrm{D}$ can break the gauge symmetry in two different ways~~\cite{Koca:1997td, Koca:2003jy,Berger:2009tt}.\footnote{$\mathrm{SU}(2)_\mathrm{D} \rightarrow {\rm U(1)}$ may appear possible by giving $\phi_0$ a VEV; but such a solution is a saddle point, and not a local minimum~\cite{Koca:2003jy}.} 
\begin{itemize}
\item $\mathrm{SU}(2)_\mathrm{D} \rightarrow T'$: For $\lambda_2 > 0$ the gauge symmetry is fully broken, but a discrete $A_4$ symmetry remains unbroken \cite{Berger:2009tt}.  In fact, the unbroken symmetry of the full Lagrangian, including the Yukawa interactions, is $T'$, the double cover of $A_4$, as the theory employs half-integer representations of fermions.  This is the desired minimum, at zero temperature, that we adopt. It is realized when $\phi_2$ acquires a nonzero VEV, and the condition $\lambda_2 > 0$ follows from the positivity of the squared masses of scalars.

\item  $\mathrm{SU}(2)_\mathrm{D} \rightarrow Q_6$: For $\lambda_2 < 0$, one realizes a $D_3$ symmetric vacuum. Including the fermion sector the unbroken symmetry would be $Q_6$, the double cover of $D_3$. This happens when $\phi_3$ acquires a nonzero VEV.  Positivity of the squared masses of scalars in this scenario requires $\lambda_2 < 0$.  If we were to adopt this chain, as can be seen from Eq. (\ref{Yuk}), two of the components of $\psi$ would remain massless, which is inconsistent with Planck data on the number of light degrees of freedom in the early universe cosmology.

\end{itemize}

It is noteworthy that for any given choice of parameters, either the $T'$ symmetry or the $Q_6$ symmetry will be realized, with no freedom to continuously go from one vacuum to the other at finite temperature.  Since at $T=0$, we have chosen the $T'$-symmetric vacuum, at finite temperature, the theory can only realize an unbroken $\mathrm{SU}(2)_\mathrm{D}$ vacuum, or the $T'$-symmetric vacuum. {\footnote By choosing $\lambda_2 > 0$ at zero temperature, we effectively disallow the $Q_6$ vacuum unless $\lambda_2$ turns negative at high temperatures due to renormalization group flow. This scenario is unlikely to occur over a short range of temperature variation, provided the Yukawa coupling of the $\Phi$ field is not excessively large.} This feature will be relevant in our discussion of phase transition in Sec. \ref{sec:phase_transition}.

For the $T^{\prime}$-symmetric vacuum, minimization of the potential at $T = 0$ gives the conditions 
\begin{align}
 \mu_H^2 =& -\lambda v^2 - \lambda_{H\Phi}v_D^2 ,\\
m^2 =& -\lambda_1  v_D^2   - \lambda_{H\Phi} v^2.
\label{eq:min}
\end{align}
where $v$ and $v_D$ are the vacuum expectation values of the fields $h$ and $s$ fields of Eq. (\ref{eq:expand}).
Under $T^{\prime}$ symmetry, the $7$-plet scalar field decomposes as $\bf {3}+\bf {3}+\bf {1}$. 
Upon spontaneous symmetry breaking one of the {\bf 3}s, denoted as $\pi_j$, will be the Goldstone bosons eaten up by the 
$\mathrm{SU}(2)_\mathrm{D}$ gauge bosons. The other {\bf 3}, denoted here as $S_j$,  is a triplet of physical scalars, while the singlet {\bf 1} will mix with the Higgs boson from the doublet.  These fields can be identified as:
\begin{align}
 &\pi_1 = \frac{\sqrt{3}}{2} \text{Re} \phi_3 + \frac{\sqrt{5}}{2} \text{Re} \phi_1, 
 &\pi_2 = -\frac{\sqrt{3}}{2} \text{Im} \phi_3 + \frac{\sqrt{5}}{2} \text{Im} \phi_1, \quad
 &\pi_3 = \sqrt{2}\text{Re}\phi_2,\\
 & S_1 = -\frac{\sqrt{5}}{2} \text{Re} \phi_3 + \frac{\sqrt{3}}{2} \text{Re} \phi_1, 
 & S_2 = -\frac{\sqrt{5}}{2} \text{Im} \phi_3 - \frac{\sqrt{3}}{2} \text{Im} \phi_1, \quad
 & S_3 = \phi_0.
\end{align}
The field-dependent masses of these scalars, needed for the computation of the effective potential for the model, are given by
\begin{align}
    &M_S^2(h,s)=m^2+\lambda_{H\phi}h^2+(\lambda_1+20\lambda_2)s^2,\\
    &M_{\pi}^2(h,s)=m^2+\lambda_{H\phi}h^2+\lambda_1s^2.
    \label{eq:msq}
\end{align}
The physical masses of the scalar triplets $S_j$ and $\pi_j$ at zero temperature can be obtained as $M^2_S=20\lambda_2 v_D^2$ and $M_\pi^2 =0$ by substituting $h= v, s=v_D$ in Eq. (\ref{eq:msq}) and using the relations of Eq. (\ref{eq:min}). The two $T^{\prime}$ singlet fields, from $\phi$ and $H$, mix with a mass-squared matrix in the basis $\left \{ \sigma,\sigma_7 \right \}$ given by (at $T=0$)
\begin{equation}
     \mathcal{M}^2_{h,h'}(h,s)=\begin{pmatrix}
   \mu_H^2+3\lambda h^2+\lambda_{H\phi}s^2  &  2\lambda_{H\phi} hs \\
   2 \lambda_{H\phi} hs &  m^2+\lambda_{H\phi}h^2+3\lambda_1s^2 
\end{pmatrix}
\end{equation}

The three massive $\mathrm{SU}(2)_\mathrm{D}$ gauge bosons form a triplet under $T^{\prime}$ symmetry with its mass given by
\begin{equation}
    M^2_V(s)=4g_D^2s^2,
\end{equation}
where $g_D$ is $\mathrm{SU}(2)_\mathrm{D}$ gauge coupling. 

The $4$-plet fermion $\psi$, after symmetry breaking, decomposes under $T'$ as  $\bf {4}=\bf {2'}+\bf {2''}$. They obtain a common mass from the Yukawa interaction term of~\autoref{Yuk},  
\begin{align}
 {\cal L}\supset i y_D s \left(\psi_{3/2} \psi_{1/2} - \psi_{-1/2} \psi_{-3/2}\right),
\end{align}
with the common  mass connecting the ${\bf 2'}$ and ${\bf 2''}$ given by
\begin{equation}
    m_\psi(s)=y_Ds.
\end{equation}
\subsection{Parameters of the model}

The parameters of the model can be collected as
\begin{align}
 \left(
  v, m_h,  \theta, m_{\psi}, m_{h'}, m_{S}, m_V, y_D
 \right),
 \label{eq:param}
\end{align}
where $m_X$ denotes the zero-temperature mass of particle $X$. 
We can express the other parameters of the model in terms of these eight input parameters:
\begin{align}
 v_D =&  \frac{m_\psi}{y_D}, \\
 g_D =&  \frac{m_V}{2 m_\psi} y_D, \label{eq:gD}\\
 \lambda =& \frac{m_h^2 \cos^2\theta + m_{h'}^2  \sin^2\theta}{2 v^2}, \label{eq:lam}\\
 \lambda_1 =& \frac{m_{h'}^2 \cos^2\theta + m_h^2 \sin^2\theta}{2 m_\psi^2} y_D^2, \label{eq:lam1}\\
 \lambda_2 =& \frac{m_S^2}{20 m_\psi^2} y_D^2, \label{eq:lam2}\\
 \lambda_{H\phi} =& \frac{(m_{h'}^2-m_h^2) \cos\theta \sin\theta}{2 v m_\psi } y_D. \label{eq:lamHPhi}
\end{align}

Two of the eight parameters listed in Eq. (\ref{eq:param}) are known, $v = 246$ GeV and $m_h = 125$ GeV, which we keep fixed.  For our numerical study, we vary the other six parameters in their allowed perturbative ranges.  These parameters are subject to various theoretical and experimental constraints, which are summarized in the next section.  In addition, we demand that the correct dark matter relic abundance $ \Omega h^2=0.1198 \pm 0.0012$~\cite{Planck:2018vyg} is generated. In practice, we first fix the parameters $(\theta,\, m_{h'})$ at specific values, and vary the other four parameters, and repeat the analysis for different fixed values of $(\theta,\, m_{h'})$.

\section{Constraints on Model Parameters}
\label{sec:constraints}
The model is constrained by various theoretical and experimental considerations. Detailed derivations of these constraints of this model are given in Ref.~\cite{Abe:2019zhx}. We briefly summarize these constraints here.  In addition to those given in Ref.~\cite{Abe:2019zhx}, we have also included the Linde-Weinberg limit~\cite{Linde:1975sw,Weinberg:1976pe} on the Higgs potential parameters arising from the condition that the broken symmetric vacuum lies deeper than the unbroken vacuum.
\subsection{Boundedness of the potential}
For the vacuum to be stable, the scalar potential has to be bounded from below along all directions in field space. The condition $\lambda> 0$ is necessary for boundedness, as can be seen by taking large values of the $H$ field with finite values for the $\phi$ fields.  $\lambda_1 >0$ is found to be necessary by going along the direction $H = 0$ along with $\phi_1=\phi_2=0$ and $|\phi_3|=\frac {2}{\sqrt 5}\phi_0$, in which case the $\lambda_2$ term gives zero contribution to the potential. The full set of necessary and sufficient conditions for a bounded potential have been derived to be:
\begin{align}
    \left \{ \lambda, \lambda_1, \tilde{\lambda}_1, \lambda_{H\phi} + \sqrt{\lambda \tilde{\lambda}_1} \right \}>0,
    \label{eq:bfb-last}
\end{align}
where
\begin{align}
   \tilde{\lambda}_1 = \lambda_1 + \min\left(0, \frac{25}{2} \lambda_2 \right).
\end{align}
\subsection{Perturbative unitarity constraints}
The scattering amplitude at high energies for two-to-two process can be expressed as
\begin{equation}
A(\alpha)=16 \pi \sum_{J}(2J+1)a_JP_J(\cos \alpha),
\end{equation}
where $J$ denotes the spin of the particle, $P_J$ stands for the Legendre polynomial and $\alpha$ is the scattering angle. Perturbative unitarity requires  the maximum eigenvalue of s-wave amplitude matrix $|Re(a_{0}^{max})| \leq \frac 12$~\cite{Lee:1977eg}. Considering  scattering processs that only contains $7-$plet scalar fields in both initial and final states, the following upper limits on the quartic couplings have been derived~\cite{Abe:2019zhx}:
\begin{align}
 8 \pi > \max\left\{ \left|2\lambda_1 - 30\lambda_2 \right|, \
 \left|2 \lambda_1 + 25 \lambda_2 \right|, \
 \left|9\lambda_1 + 60\lambda_2 \right|, \
 \left|2 \lambda_1 + 61\lambda_2 \right| \right\}.
\label{eq:PU}
\end{align}

From the scattering process $\phi\phi\rightarrow VV$, where $V$ stands for transversely polarized vector gauge bosons of $\mathrm{SU}(2)_\mathrm{D}$, an upper limit on the gauge coupling $g_D$ can be derived~\cite{Abe:2019zhx}:
\begin{equation}
|g_D| \leq 0.985.
\end{equation}

From scattering of a fermion pair into scalar boson pairs, an upper limit on the Yukawa coupling $y_D$ can be obtained: 
\begin{equation}
y_D \leq \left ( \frac {10}{\sqrt 3} \right )^{1/2}\simeq 2.4.
\end{equation}
\subsection{Linde-Weinberg limit}
For the broken symmetric vacuum to be absolutely stable, we require it to be the global minimum of the Higgs potential at zero temperature. The one-loop corrections to the potential at zero temperature involving the gauge boson sector and the Yukawa sector may change the absolute minimum of the tree-level potential.  We evaluate the Coleman-Weinberg effective potential~\cite{Coleman:1973jx}, including these contributions, and demand that the tree-level broken symmetric minimum is a deeper minimum compared to the one-loop improved unbroken symmetric vacuum~\cite{Linde:1975sw, Weinberg:1976pe}.  The details of this calculation are outlined in Sec. \ref{sec:effpot}, see Eq. (\ref{eq:CW}).  Following the prescriptions given in Ref.~\cite{Basecq:1988cv, Politzer:1978ic} we require
\begin{equation}
 V_{\mathrm{eff}}(h=v,s=v_D) \leq 0.
\label{eq:WL}
\end{equation}
We implement this condition numerically in our analysis.  
In the $\lambda_{H \phi} =0$ limit, the above constraint can be written analytically as
\begin{align}
    m^2_{h^\prime} v_D^2 \leq & \frac {v_D^4}{16\pi^2}\left [ 12\lambda_1^2+600\lambda_2^2+120\lambda_1\lambda_2+144g_D^2-8y_D^2 \right ] \nonumber\\
    &+\frac {(\lambda_1v_D^2)^2}{8\pi^2}\log\left ( \frac {2\lambda_1v_D^2}{m_{\psi}^2} \right )+\frac {3(20\lambda_2v_D^2)^2}{64\pi^2}\log\left ( \frac {20\lambda_2v_D^2}{m_{\psi}^2} \right )~.
\end{align}
We note that this condition was not implemented in the analysis of Ref. \cite{Abe:2019zhx}.
\subsection{Higgs boson observables}
There are limits on the mixing of the SM-like Higgs with the singlet Higgs ($h-h'$ mixing) parametrized by the angle $\theta$ from Higgs observables.  
In the presence of such mixing, the coupling of the SM-like Higgs is modified from the SM Higgs coupling by a factor $\cos\theta$:
    \begin{equation}
        \frac {g_{WWh}}{g_{WWh}^{SM}}=\frac {g_{ZZh}}{g_{ZZh}^{SM}}=\frac {g_{ffh}}{g_{ffh}^{SM}}=\cos \theta,
    \end{equation}
 where $f$ denotes any SM fermion. For scalar $h^\prime$ in the mass range $\left \{ 250,1000 \right \}\,\mathrm{GeV}$, $W$-boson mass correction provides an upper limit  $\left | \theta \right |<0.22$~\cite{Robens:2016xkb,ATLAS:2015ciy}. For $m_{h^\prime}<250\, \mathrm{GeV}$, the upper limit for mixing angle comes from direct Higgs searches and the observed Higgs signal strength at colliders, with ATLAS~\cite{ATLAS:2016neq} experiment providing an upper limit $\left | \theta \right |\lesssim 0.25$.
\subsection{Dark matter constraints}
The four components of $\psi$, which transform as ${\bf 2'} + {\bf 2''}$ of the unbroken $T'$ symmetry, are absolutely stable and are dark matter candidates. Since $T'$ is unbroken, the lightest $T'$-nonsinglet in the model must be stable.  This may be the $\psi$ fermion, or the triplet scalar $H$, or the triplet vector gauge boson $V^\mu$.  In the last two cases, the theory would have two stable dark matter particles, each contributing to the relic density, since the fermion is always a DM candidate owing to fermion number conservation. In Ref.~\cite{Abe:2019zhx}, all these cases were analyzed, including the possibility of multi-component DM.  In the present analysis, we confine to the scenario where the fermionic $\psi$ is the lightest $T'$-nonsinglet particle.

In this work, we are primarily interested in the single-component dark matter scenario, which exhibits promising GW signals.
We assume DM is thermally produced in the early Universe, and DM particles were in the thermal bath made by the SM particles. As the Universe expands, $\psi$ decouples from the thermal bath, and the number density is determined by the freeze-out mechanism~\cite{Lee:1977eg}. The evolution of the number density of the DM $n$ is determined by the Boltzmann equation,
\begin{align}
\dv{n}{t} + 3 H n = - \expval{\sigma v} \qty( n^2 - n^{eq.} ),
\end{align}
where $H$ is the Hubble parameter, $\expval{\sigma v}$ is the thermal average of $\sigma v$ for DM annihilation processes, and $n^{eq.}$ is the number density of DM in the thermal bath. 
To explain the measured value of the DM energy density $\Omega h^2 = 0.120 \pm 0.001$~\cite{Planck:2018vyg}, the canonical value of $\expval{\sigma v} \simeq 3 \times 10^{-26} \text{ cm}^3 \text{s}^{-1}$ is required. 
To obtain the canonical value of $\expval{\sigma v}$, we tune a model parameter, which we choose $y_\text{DM}$ in our analysis. In our setup, the main annihilation processes are $\psi \bar{\psi} \to f \bar{f}, hh, hh'$, and $h'h'$, where $f$ stands for the SM fermions~\cite{Abe:2019zhx}. The dominant annihilation process is $\psi \psi \rightarrow h^\prime h^\prime$, as the relevant Yukawa coupling is not suppressed by the small mixing angle $\theta$. If this channel is not kinematically open, the dominant channel would be $\psi \psi \rightarrow b \overline{b}$, which would require a large value for $y_D$, as this coupling has a suppression factor $\theta$.  The pair-annihilation process of DM into $h^\prime h^\prime$ is efficient for $m_{\psi}\gtrsim m_{h^\prime}$, and a large value of $y_D$ is not required. 

We also consider the constraints from DM direct detection experiments such as XENONnT~\cite{XENON:2023cxc} and LZ~\cite{LZ:2022lsv} experiments. They give upper bounds on the WIMP-nucleon scattering cross-section. In our analysis, $\psi$ scatter off a nucleon via exchange of $h$ and $h'$ in $t$-channel. The spin-independent scattering cross section $\sigma_\text{SI}$ is given by~\cite{Abe:2019zhx},
\begin{align}
  \sigma_\text{SI} 
  = \sin^2 2 \theta 
  \qty( \frac{1}{m_h^2} - \frac{1}{m_{h'}^2} )^2
  f_N^2 \frac{m_\psi^4 m_N^4}{v_D^2 v^2 (m_\psi + m_N)^2 4\pi},
\end{align}
where $f_N \simeq 0.3$.  The LZ experiment gives the stringent bound. It gives $\sigma_\text{SI} < 9.2\times 10^{-48} \text{ cm}^2$ for $m_\text{DM} = 36$~GeV.
In the following analysis, we choose benchmark parameter points that satisfy the constraints on $\sigma_\text{SI}$. Typically, it requires a very small mixing angle $|\sin\theta|\ll 1$.

We use \texttt{microOMEGAs}~\cite{Belanger:2018ccd} to calculate DM relic abundance and WIMP-nucleon scattering cross-section. In the following section, we evaluate the GW signals of the parameter points that give the measured value of the DM energy density $\Omega h^2  = 0.12$ and evade the constraints from the DM direct detection experiments.

\section{Finite-Temperature Effective Potential}
\label{sec:effpot}

Finite temperature effects determine the dynamics of cosmological phase transitions and, consequently, spontaneous symmetry breaking in the early universe. These effects are described by the effective potential at finite temperatures. Here we present a full one-loop calculation of the finite temperature effective potential for the model.  This perturbative method, while simple and straightforward, does have some limitations, such as residual gauge dependence, which won't be present in a full non-perturbative lattice calculation, which is, however, more challenging to carry through. In our one-loop analysis, we have included the ring diagrams (or the daisy diagrams), which can be important due to their infrared divergent behavior.

In our analysis we are interested in $\lambda_2>0$, in which case $\mathrm{SU}(2)_\mathrm{D}$ breaks down to $T^\prime$ and not $Q_6$ at $T=0$.  The one-loop effective potential is given by\footnote{The effective potential exhibits theoretical uncertainties owing to the choice of gauge parameter~\cite{Patel:2011th,Arunasalam:2021zrs,Chiang:2017nmu,Garny:2012cg,Metaxas:1995ab,Wainwright:2011qy}. The evaluation of these uncertainties is an active field of research; for recent developments, see for example Ref.~\cite{Gould:2019qek,Croon:2020cgk,Gould:2021oba,Niemi:2021qvp,Schicho:2021gca}.}
\begin{equation}
V_{\mathrm{eff}}(h,s,T)=V_0(h,s)+V_{\mathrm{CW}}(h,s)+V_{\mathrm{1T}}(h,s,T)+V_{\mathrm{daisy}}(h,s,T)+V_{\mathrm{CT}}(h,s).
\end{equation}
Here $V_0$ is the tree-level potential  expressed in terms of classical field values $(h,s)$ and is given by
\begin{equation}
    V_{0}(h,s)=\frac{\mu_H^2}{2}h^2+\frac{\lambda}{4}h^4+\frac{m^2}{2}s^2+\frac{\lambda_{H\phi }}{2}h^2s^2+\frac{\lambda_1}{4}s^4.
\end{equation}

The Coleman-Weinberg term $V_{\mathrm{CW}}$~\cite{Coleman:1973jx} describes zero temperature one-loop corrections to the potential. Using $\overline{\mathrm{MS}}$ renormalization procedure, this term  can be written in the Landau gauge as
\begin{equation}
    V_{\mathrm{CW}}=\sum_i{\frac{n_i}{64\pi^2}}M^4_i(h,s)\left [ \log\left ( \frac {M^2_i(h,s)}{\mu^2} \right ) -c_i\right ],
    \label{eq:CW}
\end{equation}
where the index $i\in \left \{ h^\prime,h,S_j, \pi_j,G,V,W^\pm,Z,\Psi,t  \right \}$ runs over all particles in the thermal bath, $n_i= \left \{ 1,1,3,3,3,9,6,3,-8,-12 \right \} $ denotes the number of degrees of freedom for the particles with $n_i>0$ ($n_i<0$) for bosons (fermions), and $c_i = \frac{3}{2} \left( \frac{5}{6} \right)$ for scalars and fermions (gauge bosons). Here, we have ignored the masses of all SM fermions, except for the top quark.

We choose the renormalization scale $\mu^2=m_{\psi}^2$.\footnote{Regarding the dependence of renormalization scale $\mu^2$ of the effective potential at finite temperature, we refer to Ref.~\cite{Gould:2021oba}.} The masses and mixing angle in the model extracted from the one-loop correction at $T=0$, will have a small variation compared to the values extracted from tree-level potential. For a systematic scan and consistent DM analysis, it is convenient to use the one-loop corrected mass and mixing angle to be the same as the the tree-level values which are used as input. This can be achieved by adding finite terms in counterterm potential $V_{\mathrm{CT}}$ using the renormalization scheme given in Ref.~\cite{Basler:2016obg}. The counterterms can be written as
\begin{equation}
    V_{\mathrm{CT}}(h,s)=\frac{\delta \mu_H^2}{2}h^2+\frac{\delta \lambda}{4}h^4+\frac{\delta m^2}{2}s^2+\frac{\delta \lambda_{H\phi }}{2}h^2s^2+\frac{\delta \lambda_1}{4}s^4.
\end{equation}

We choose the following renormalization conditions:
\begin{align}
     &
    \frac{\partial V_{\mathrm{CT}}(h,s) }{\partial x_i}\bigg\rvert_{h=v,s=v_D}=-\frac{\partial V_{\mathrm{CW}}(h,s) }{\partial x_i}\bigg\rvert_{(h=v,s=v_D)},\\
    &
    \frac{\partial^2 V_{\mathrm{CT}}(h,s) }{\partial x_i \partial x_j}\bigg\rvert_{h=v,s=v_D}=-\frac{\partial^2 V_{\mathrm{CW}}(h,s) }{\partial x_i \partial x_j}\bigg\rvert_{(h=v,s=v_D)},
    \label{eq:CTd2}
\end{align}
for $x_i \in \{h,s \} $. We can use these five independent conditions to determine the five parameters of the counterterms. In this case, we can use the tree-level relations derived for the various parameters, see Eq. (\ref{eq:lamHPhi}). The second derivative of $V_{\mathrm{CW}}$ in~\autoref{eq:CTd2} leads to infrared divergences when including contribution from the Goldstone boson in Landau gauge~\cite{Cline:1996mga,Casas:1994us, Elias-Miro:2014pca,Camargo-Molina:2016moz,Cline:2011mm,Dorsch:2013wja}. 
In the complete theory, this IR divergence of the effective potential evaluated in the zero-momentum limit will cancel with momentum-dependent contributions to the vertices~\cite{Casas:1994us, Elias-Miro:2014pca}. In this work, we use the prescription given in Ref.~\cite{Camargo-Molina:2016moz}, where an IR regulator mass is added for the Goldstone boson to identify and then discard the IR divergence. 

The one-loop finite temperature corrections to the potential are given by~\cite{Dolan:1973qd,Quiros:1999jp}
\begin{equation}
V_{\mathrm{1T}}(h,s,T)=\frac {T^4}{2\pi^2}\sum_i J_i\left ( \frac {M_i(h,s,T)^2}{T^2} \right ),
\end{equation}
where the index $i$ runs over all scalars, vector bosons (both transverse and longitudinal components), and fermions. The thermal functions $J_{B/F}(x)$ are given by the integrals
\begin{equation}
J_{B/F}(y^2)=\int_0^\infty dx x^2\log(1\mp e^{-\sqrt{x^2+y^2}}).
\end{equation}

We have also included the ring diagram contributions (or the Daisy contributions)~\cite{Carrington:1991hz, Arnold:1992rz}, which turn out to be important owing to their infrared behavior. This class of diagrams are not captured in the mean-field approximation used in the derivation of the effective potential.  They can be resummed and lead to the following corrections~\cite{Arnold:1992rz}:
\begin{equation}
   V_{\mathrm{daisy}}(h,s,T)=-\frac {T}{12\pi}\left [ \sum_{i}\left ( (\bar{M_i}^2)^{\frac 32}-(M_i^2)^{\frac 32} \right ) \right ].
\end{equation}
Here $\bar{M_i}$ is the thermal Debye mass of the scalar species $i$ or the longitudinal components of gauge boson. For scalars $\bar{M_i}$ is evaluated with the replacement $\mu^2_{H} \to \mu^2_{H}+ \Pi_h T^2$ and $m^2 \to m^2+ \Pi_s T^2$, where $\Pi_h$ and $\Pi_s$ are given by 
\begin{align}
    \Pi_h=&\frac {\lambda_{H\phi}}{12}\left ( \theta_{h^\prime}+3\theta_{H}+3\theta_{\Pi_H} \right )+\frac{\lambda}{4}\left ( \theta_{h}+\theta_{\Pi_h} \right )+\frac {y_t^2}{2}\theta_t+\frac{g^2}{16}\left (2\theta_{W^\pm}+\theta_{W^3} \right)+\frac{g'^2}{16}\theta_B, \\
    \Pi_s=& \frac {\lambda_1}{4}\left ( \theta_{h^\prime}+\theta_{H}+\theta_{\Pi_H} \right )+5\lambda_2 \theta_H+3g_D^2\theta_{V}+\frac {y_D^2}{3}\theta_{\Psi}+\frac {\lambda_{H\phi}}{12}\left ( \theta_{h}+3\theta_{\Pi_h} \right ).
\end{align}
Here $\theta_i$ are step functions defined as follows. We only consider the contribution from a particle species $i$ to $V_{\rm eff}$ with mass $M_i<T$~\cite{Comelli:1996vm,Katz:2014bha} and  $\theta_i=\theta(T^2-M_i^2)$ is the step-function, equal to 1 if $T^2>M_i^2$ and 0 if $T^2<M_i^2$.
For the longintudnal components of gauge bosons $\bar{M_i}$ are given by, $\bar{M^2_i}=M^2_i+\Pi_iT^2$. For the $\mathrm{SU}(2)_\mathrm{D}$ gauge bosons $\Pi_V$ is given by
\begin{equation}
    \Pi_V=g_D^2\left ( \frac 23 \theta_{V}+\frac 13 \left ( 4\theta_{h^\prime}+12\theta_{H}+12\theta_{\Pi_H} \right ) +\frac 56 \theta_{\Psi}\right ).
\end{equation}
The thermal masses for standard model gauge bosons are given, for e.g., in Ref.~\cite{Katz:2014bha}.

\section{Phase Transition}
\label{sec:phase_transition}
Using the finite temperature effective potential described in the previous section, we can study the thermal evolution of the universe for the model. At a very high temperature, the global minimum of the potential is at the origin  $(h=s=0)$, the symmetric vacuum where the SM gauge symmetry as well as the $\mathrm{SU}(2)_\mathrm{D}$ gauge symmetry are unbroken. As the universe cools down, additional minima begin to appear. We use publicly available Python code \texttt{CosmoTransitions}~\cite{Wainwright:2011kj} to find these minima and trace their thermal evolution. If two minima coexist over a temperature range, separated by a barrier, a first-order phase transition can occur.

The model realizes a two-step phase transition. In one step, we have electroweak symmetry breaking proceeding as in the SM since the cross-couplings $\lambda_{H\phi}$ that connects the dark sector and the SM sector is very small.\footnote{Smallness of $\lambda_{H\phi}$ is needed for consistency with direct detection limits for most of the parameter space.}
In the other step, $SU(2)_D\rightarrow T'$ symmetry breaking occurs, which is a mostly first-order phase transition. Detectable GW signals in the model are realized mainly when electroweak phase transition occurs first, followed by breaking of $\mathrm{SU}(2)_\mathrm{D}$ symmetry:
\begin{equation}
    (h=0,s=0)\xrightarrow[]{\text{cross-over}}(h\neq0,s=0)\xrightarrow[]{\text{first-order}}(h\neq0,s\neq0).
\end{equation}
\begin{figure}[tb]
\centering
\includegraphics[width=0.49\hsize]{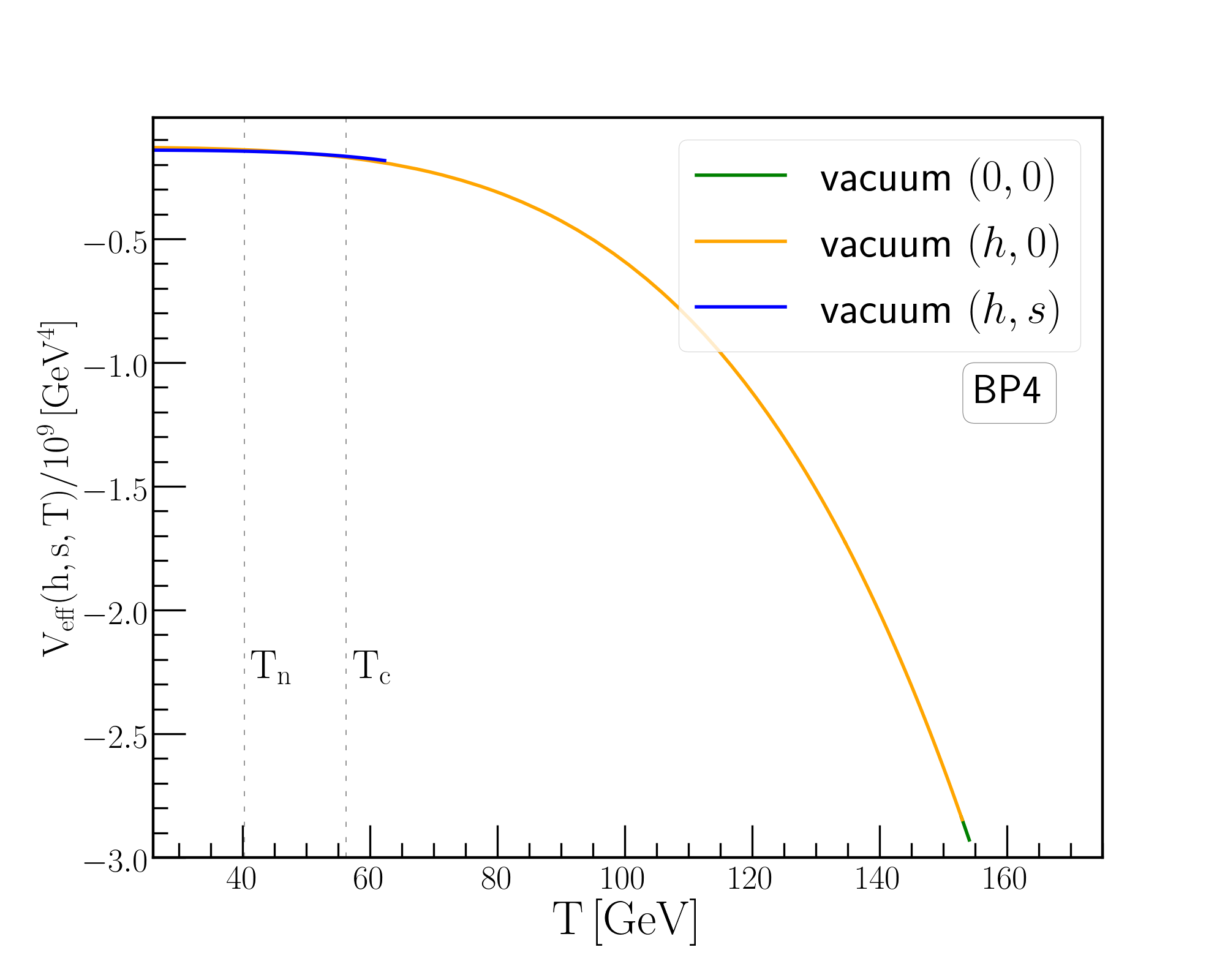}
\includegraphics[width=0.49\hsize]{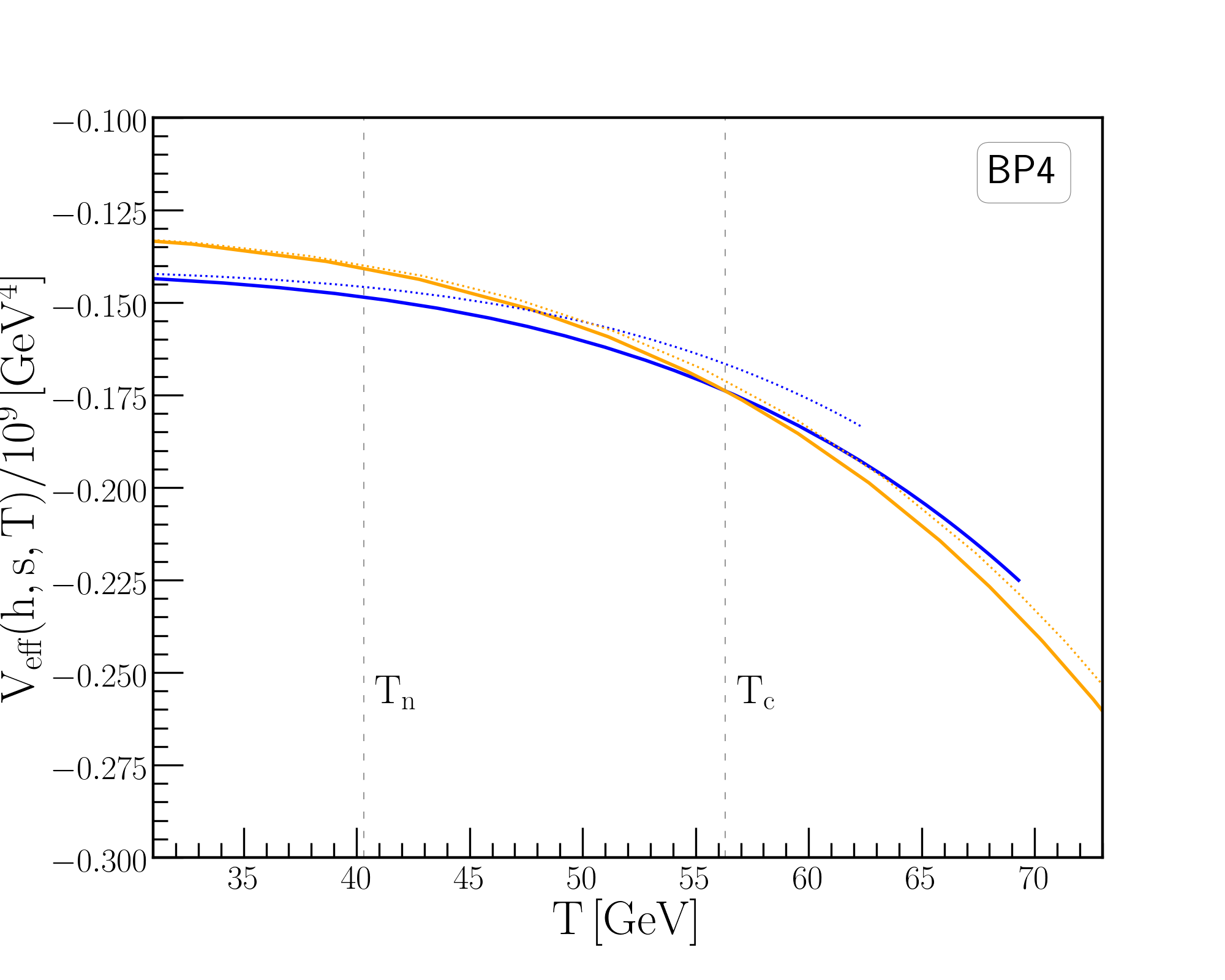}
\label{fig:running}
\caption{Evolution of effective potential $ V_{\mathrm{eff}}(h,s,T)$ with temperature for different vacua for the benchmark point BP4 shown in Table \ref{tab:1} . A second-order phase transition occurs along the $h$ axis first, followed by a first-order phase transition along the $s$ axis. The black dotted lines denote the critical temperature and nucleation temperature for the first-order phase transition. The temperature range in which $\mathrm{SU}(2)_\mathrm{D}$ breaking occurs is zoomed-in on the left panel. The dotted line shows effective potential contribution without including the daisy terms.}
\end{figure}

For the first-order phase in the second stage, at the transition temperature at $T_c$, the true vacuum and false vacuum are degenerate. At temperatures below $T_c$, the true vacuum becomes the global minimum and tunneling from the false vacuum to the true vacuum can occur in the form of bubble nucleation. At finite temperature, the bubble nucleation per unit time and unit volume is given by~\cite{Linde:1980tt,Coleman:1977py}
\begin{equation}
    \Gamma (T)\approx T^4\left (\frac {S_3}{2\pi T}  \right )^{3/2}e^{-\frac {S_3}{T}},
\end{equation}
where $S_3$ is the three-dimensional Euclidean action corresponding to the critical bubble. We can calculate $S_3$ using
\begin{equation}
   S_3(h,s,T)=\int_{0}^{\infty}{drr^2\left [ \frac 12\left ( \frac {dh}{dr} \right )^2+\frac 12\left ( \frac {ds}{dr} \right )^2+ V_{\mathrm{eff}}(h,s,T) \right ]},
\end{equation}
where $r$ is the radius of the path. Assuming spherical symmetry, the tunneling path can be found by minimizing $S_3$ and thus solving the differential equations,
\begin{align}
    &\frac {d^2h}{dr^2}+\frac 2r \frac{dh}{dr}=\frac {d V_{\mathrm{eff}}(h,s,T)}{dr},\\
    &\frac {d^2s}{dr^2}+\frac 2r \frac{ds}{dr}=\frac {d V_{\mathrm{eff}}(h,s,T)}{dr},
\end{align}
subject to the boundary conditions
\begin{align}
    \frac {dh}{dr}\Bigr|_{r=0}=\frac {ds}{dr}\Bigr|_{r=0}=0
    \hspace{0.5cm} \text{and} \hspace{0.5cm}
      (h,s)\Bigr|_{r \rightarrow \infty} \equiv \text{False vacuum}.
\end{align}

Here the tunneling path with $r=0$ corresponds to a point near the true vacuum and $r=\infty$ corresponds to the false vacuum. We use  CosmoTransitions~\cite{Wainwright:2011kj} to solve these differential equations and compute $S_3$.  Bubble nucleation occurs when the tunneling rate is comparable to $H^4$, the Hubble rate per unit Hubble volume. The nucleation temperature $T_n$~\cite{Moreno:1998bq} is defined as the temperature where one bubble nucleates per horizon volume,
\begin{equation} 
\label{equ:Tncond}
    \int_{T_n}^{\infty}\frac {dT}{T}\frac {\Gamma (T)}{H(T)^4}=1.
\end{equation}
We can approximate~\autoref{equ:Tncond} as $\Gamma(T)/H(T)^4\sim 1$ and assuming temperature $\mathrm{O}(100\,\mathrm{GeV})$ the above condition can be written as $\frac {S_3(T)}{T}\approx 140$ ~\cite{Quiros:1999jp}. { As the phase transition progresses, true vacuum bubbles begin to emerge and expand within the false vacuum, and the fraction of false vacuum is provided by~\cite{Ellis:2018mja, PhysRevD.46.2384}
\begin{equation}
F(T)=\mathrm{exp}\left [ -T^3\int_T^{T_c}\frac {dT^\prime\Gamma(T^\prime)}{{T^\prime}^3 H(T^\prime)^3}\left (\int_T^{T^\prime}\frac {d\tilde{T}}{H(\tilde{T})}  \right )^3 \right ].
\end{equation}
The percolation temperature is defined as the temperature in which $F(T)=0.71$ of the universe is covered by the false vacuum, and the existence of percolation temperature ensures the completion of phase transition~\cite{Badger:2022nwo,Athron:2022mmm}. The supercooling phase transition occurs when the nucleation temperature is considerably smaller than the nucleation temperature, which leads to $\alpha \gg 1$. In this model, we have all the points with $\alpha<1$, and we assume there is no supercooling and the percolation temperature can be approximated to the nucleation temperature $T_p\simeq T_n$. To demonstrate this approximation, we have computed the percolation temperature explicitly for the four benchmark points provided in~\autoref{tab:1}.} Here, we consider $T_n$ as the gravitational wave production temperature. At the nucleation temperature, we can define two important phenomenological quantities for estimating the GW spectrum. One is $\alpha$, a measure of phase transition strength, which corresponds to the latent heat released during phase transition normalized by the energy density of plasma at the time of phase transition~\cite{Espinosa:2010hh},
\begin{equation}
    \alpha=\frac{\rho_{\mathrm{vac}}}{\rho_{\mathrm{rad}}}=\frac{1}{\rho_{\mathrm{rad}}}\left [\frac {T}{4} \frac {d\Delta V}{dT}-\Delta V \right ]_{T_n}.
\end{equation}
The energy density in radiation,  $\rho_{\mathrm{rad}}$, is given by $\rho_{\mathrm{rad}}=\frac {g_{*}\pi ^2T^4}{30}$, where $g_{*}=126.75$ in the present model when all species are active. The second parameter $\beta$ corresponds to the inverse time duration of phase transition~\cite{Kamionkowski:1993fg},
\begin{equation}
    \beta=\left (HT\frac {d(s_3/T)}{dT}  \right )_{T_n},
\end{equation}
where $H$ is the Hubble rate. We use $H_{\star}$ to denote the Hubble rate at $T_n$.

In Fig. 1, we show the evolution of the effective potential for the various vacua versus temperature for the benchmark point BP4 presented in Table \ref{tab:1}. 
The $\mathrm{SU}(2)_{\mathrm{D}}$ broken vacuum starts to appear around $T = 65\, \mathrm{GeV}$, the two vacua becomes degenerate around $T = 55\,\mathrm{GeV}$ and the nucleation temperature is $T_n = 40.32\,\mathrm{GeV}$. In the right panel, we show the effective potential without the daisy term contribution with dotted lines. This illustrates the importance of including the daisy terms in the analysis; without the daisy terms, the critical temperature $T_c$ would be $50.62\,\mathrm{GeV}$ instead of $56.31\,\mathrm{GeV}$.

{ In the model with our benchmark points, freeze-out occurs at a temperature scale of approximately $T \approx m_{\mathrm{DM}}/20 \simeq 5$~GeV, which is significantly lower than the phase transition temperature. Therefore, the phase transition dynamics do not significantly affect the mechanisms governing dark matter. Additionally, the thermal mass around the freeze-out temperature of $\sim5~\mathrm{GeV}$ remains negligible and does not significantly influence dark matter physics.
}

\section{Gravitational Wave Signatures}
\label{sec:GW}
First-order phase transition in the early universe could exhibit detectable GW signals today; see Ref.~\cite{Caprini:2015zlo, Cai:2017cbj,Caprini:2018mtu, Athron:2023xlk,Kierkla:2022odc,Lewicki:2022pdb,Ellis:2020nnr} for recent reviews. The GW signals in this model peak around the millihertz-hertz range, and it is possible to detect these signals with next-generation space-based detectors such as LISA~\cite{Caprini:2015zlo,Caprini:2019egz}, BBO~\cite{Corbin:2005ny} and DECIGO~\cite{Kudoh:2005as}. There are three mechanisms to generate GWs during a FOPT: bubble collision, sound waves, and magnetohydrodynamic (MHD) turbulence. The GW spectrum, denoted by $\Omega_{\mathrm{GW}}(f)$, represents the present GW energy density per logarithmic frequency interval per critical energy density of the universe.  The GW energy density can be approximated linearly as
\begin{equation}   \Omega_{\mathrm{GW}}h^2=\Omega_{\mathrm{coll}}h^2+\Omega_{\mathrm{SW}}h^2+\Omega _{\mathrm{turb}}h^2,
\end{equation}
where $h=H_0/100 \text{km}/ \text{s}/\text{Mpc}$ is the current Hubble parameter.

\subsection{Sound waves}
The energy produced during the phase transition into the plasma can either heat up the plasma or release kinetic energy, which in turn creates fluid motion. The numerical estimates indicate the fluid's energy-momentum tensor following a bubble collision resembles that of an ensemble of sound waves~\cite{Hindmarsh:2017gnf}. These sound waves have the potential to be a source of gravitational waves. The GW spectrum from sound waves is given by~\cite{Hindmarsh:2013xza, Hindmarsh:2017gnf, Hindmarsh:2016lnk}
\begin{equation}
     \Omega_{\mathrm{SW}}h^2=\frac {2.65 \times 10^{-6}}{\beta/H_{\star}}\left ( \frac {k_s \alpha}{1+\alpha} \right )^2\left ( \frac {100}{g_{*}} \right )^{1/3}v_w \left ( \frac {f}{f_{\mathrm{SW}}} \right )^3\left ( \frac {7}{4+3(f/f_{\mathrm{SW}})^2} \right )^{7/2}\Upsilon (\tau_{\mathrm{SW}}).
\end{equation}
Here $k_s$ is an efficiency factor for the sound wave spectrum, given by~\cite{Caprini:2015zlo}
\begin{equation}
    k_s=\frac {\alpha}{0.73+0.083\sqrt{\alpha}+\alpha}.
\end{equation} 

In a radiation-dominant universe, the sound waves have a finite lifetime. Due to the finite active period $\tau_{\mathrm{SW}}$ of sound waves, recent studies~\cite{Guo:2020grp, Hindmarsh:2020hop} have shown a suppression factor $\Upsilon (\tau_{\mathrm{SW}})$, given by
\begin{equation}
    \Upsilon (\tau_{\mathrm{SW}})=1-\frac {1}{\sqrt{1+2\tau_{\mathrm{SW}} H_{\star}}}.
\end{equation}

The sound wave lasts until turbulence begins to develop, and the $\tau_{\mathrm{SW}}$ is provided by~\cite{Hindmarsh:2017gnf}
\begin{equation}
    \tau_{\mathrm{SW}}=\frac {R_{\star}}{\overline{U_f}},
\end{equation}
where the mean bubble separation is denoted by $R_{\star}\simeq (8\pi)^{1/3}v_w/\beta$, while $\overline{U_f}^2$ is the mean square velocity given by~\cite{Bodeker:2017cim}
\begin{equation}
    {\overline{U_f}}^2=\frac 34 \frac {\alpha}{1+\alpha}k_s.
\end{equation}

The peak frequency of the GW spectrum from sound waves accounting for the redshift factor is given by~\cite{Huber:2008hg}
\begin{equation}
    f_{\mathrm{SW}}=\frac {1.9\times 10^{-5}}{v_w}\left ( \frac {\beta}{H_{\star}} \right )\left ( \frac {T_n}{100\mathrm{GeV}} \right )\left ( \frac {g_{*}}{100} \right )^{1/6}\text {Hz}.
\end{equation}
\subsection{Bubble collisions}
When the nucleated bubbles collide with each other, the spherical symmetry is broken, and gravitational waves are generated. In the $\frac {\beta}{H_{\star}}\gg 1$ limit, one can make the envelope approximation~\cite{Kosowsky:1991ua}, wherein the gravitational wave from the overlapped region is neglected. The gravitational wave spectrum produced by bubble collision can be approximated as~\cite{Jinno:2016vai}
\begin{equation}
     \Omega_{\mathrm{coll}}h^2=\frac {1.67 \times 10^{-5}}{(\beta/{H_{\star}})^2}\left ( \frac {k_c \alpha}{1+\alpha} \right )^{2}\left ( \frac {100}{g_{*}} \right )^{1/3}\left ( \frac {0.11v_w^3}{0.42+v_w^2} \right )\frac {3.8(f/f_{env})^{2.8}}{1+2.8(f/f_{env})^{3.8}}.
\end{equation}

The peak frequency of the GW spectrum accounting for the redshift of the frequency from bubble collision is given by~\cite{Huber:2008hg}
\begin{equation}
   f_{\mathrm{env}}=\left ( \frac {0.62 \times \beta/H_{\star}}{1.8-0.1v_w+v_w^2} \right )16.5\times 10^{-6}\left ( \frac {T_n}{100\mathrm{GeV}} \right )\left ( \frac {g_{*}}{100} \right )^{1/6} \text{Hz}.
\label{eq:peakfreqbub}
\end{equation}

The first term in~\autoref{eq:peakfreqbub} on the right side denotes the peak frequency of the GW spectrum from bubble collision at $T_n$. The  efficiency factor for the bubble collision $k_c$ is given by~\cite{Kamionkowski:1993fg}
\begin{equation}
    k_c=\frac {0.715\alpha+\frac {4}{27}\sqrt{\frac{3\alpha}{2}}}{1+0.715\alpha}.
\end{equation}

\subsection{MHD turbulence}
If the early-universe plasma has an extremely high Reynolds number, the energy injected into it can create turbulence in the fluid~\cite{Kamionkowski:1993fg}. The turbulent motion in the early-universe plasma can be a source for the gravitational waves~\cite{Witten:1984rs}. Furthermore, a fully ionized plasma fluid can give rise to a turbulent magnetic field under turbulent motion, generating gravitational waves. The GW spectrum arising from the turbulence can be modeled as~\cite{Caprini:2009yp}
\begin{equation}
   \Omega _{\mathrm{turb}}h^2=\frac {3.35 \times 10^{-4}v_w}{\beta/H_{\star}}\left ( \frac {k_t \alpha}{1+\alpha} \right )^{3/2}\left ( \frac {100}{g_{*}} \right )^{1/3} \left ( \frac {f}{f_{\mathrm{SW}}} \right )^3\frac {1}{\left [1+(f/f _{\mathrm{turb}})  \right ]^{11/3}(1+8\pi f/h_{\star})}.
   \label{eq:6.11}
\end{equation}
We choose  $k _{\mathrm{turb}}=0.05k_s$ based on suggestion from numerical stimulation~\cite{Caprini:2015zlo}. The peak frequency for GW spectrum arising from MHD turbulence is provided by~\cite{Caprini:2009yp}
\begin{equation}
    f _{\mathrm{turb}}=\frac {2.7\times 10^{-5}}{v_w}\left ( \frac {\beta}{H_{\star}} \right )\left ( \frac {T_n}{100\mathrm{GeV}} \right )\left ( \frac {g_{*}}{100} \right )^{1/6} \text{Hz}.
\end{equation}
The factor $h_{\star}$ in Eq. (\ref{eq:6.11}) accounts for the redshift of the frequency,
\begin{equation}
    h_{\star}=16.5\times 10^{-6}\left ( \frac {T_n}{100\mathrm{GeV}} \right )\left ( \frac {g_{*}}{100} \right )^{1/6} \mathrm {Hz}.
\end{equation}

{ We assume that the phase transition occurs in the detonation regime, and, we fix $v_w$ as Chapman-Jouguet velocity~\cite{PhysRevD.25.2074} which provides a lower bound for bubble wall velocity~\cite{Huber:2008hg,Addazi:2023ftv},
\begin{equation}
v_w=\frac {1+\sqrt{3 \alpha(1+c^2_{w}(3\alpha-1))}}{c^{-1}_{w}+3\alpha c_{w}},
\end{equation}
where $c_w=1/\sqrt{3}$ is the speed of sound at false vacuum.
} To provide information about the detectability of the GW signal in the detector, we define the signal-to-noise ratio, SNR~\cite{Caprini:2015zlo}, as
\begin{equation}
    \text {SNR}=\sqrt {\tau \int_{f _{\mathrm{min}}}^{f _{\mathrm{max}}}{df\left [ \frac {\Omega_{\mathrm{GW}}(f)h^2}{\Omega_{\mathrm{sens}}(f)h^2} \right ]^2}}.
\end{equation}

\section{Numerical results}
\label{sec:numer}

In our numerical study, we first scan for points in the parameter space that give the correct dark matter relic abundance while satisfying all the constraints listed in Sec. \ref{sec:constraints}.  Then, we 
analyze the phase transition pattern for these parameter points. We selected the parameter points that exhibit strong first-order phase transition ($\xi_C=\frac{s_c}{T_c}>1$), and for these points, we evaluated the parameters $\alpha$ and $\beta/{H_\star}$ at the nucleation temperature $T_n$ with the package CosmoTransitions~\cite{Wainwright:2011kj}. Using these three parameters at nucleation temperature, we compute GW spectra and evaluate SNR to gauge its accessibility to future experiments.

\begin{figure}
\includegraphics[width=0.48\hsize]{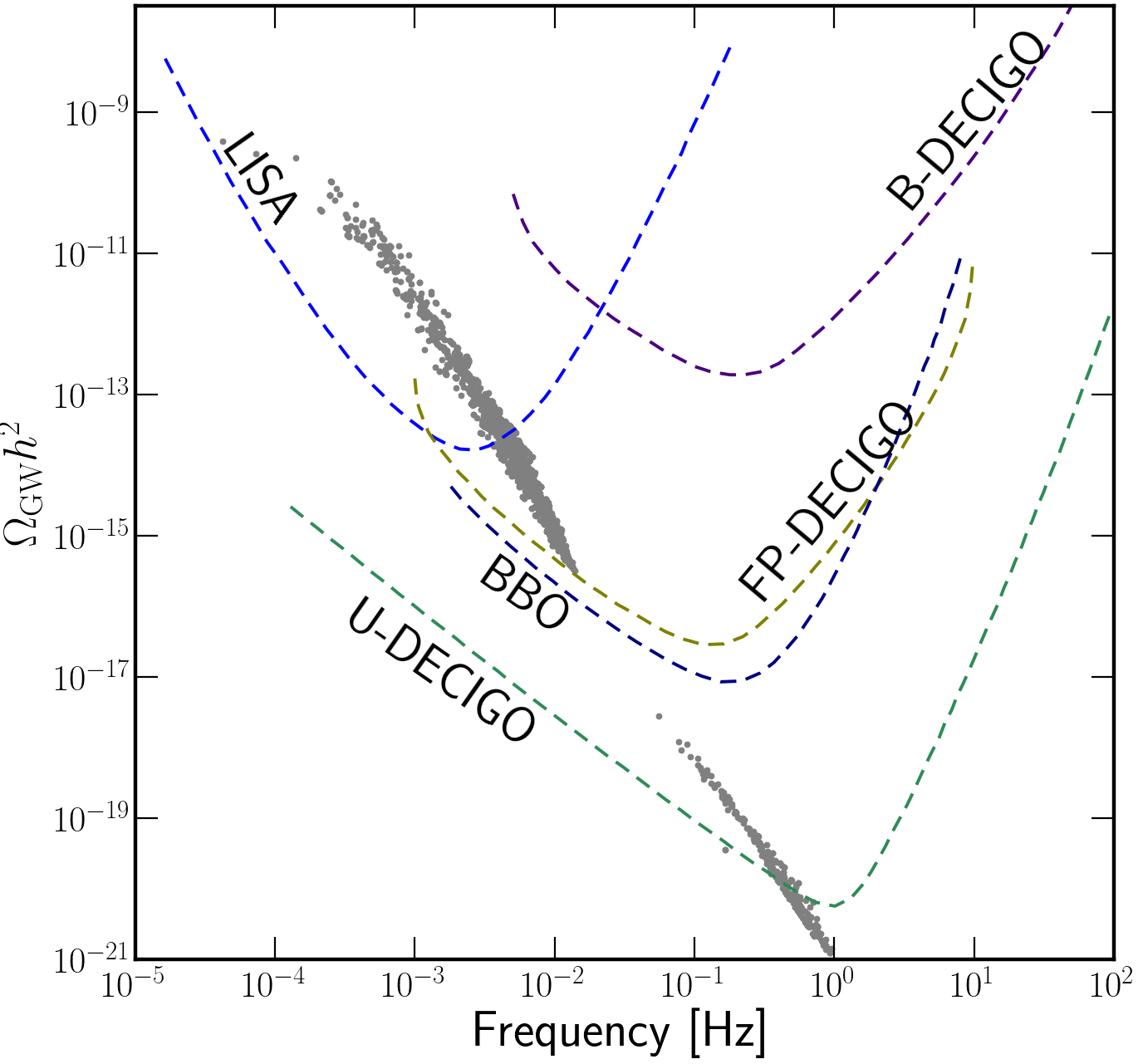} 
\includegraphics[width=0.48\hsize]{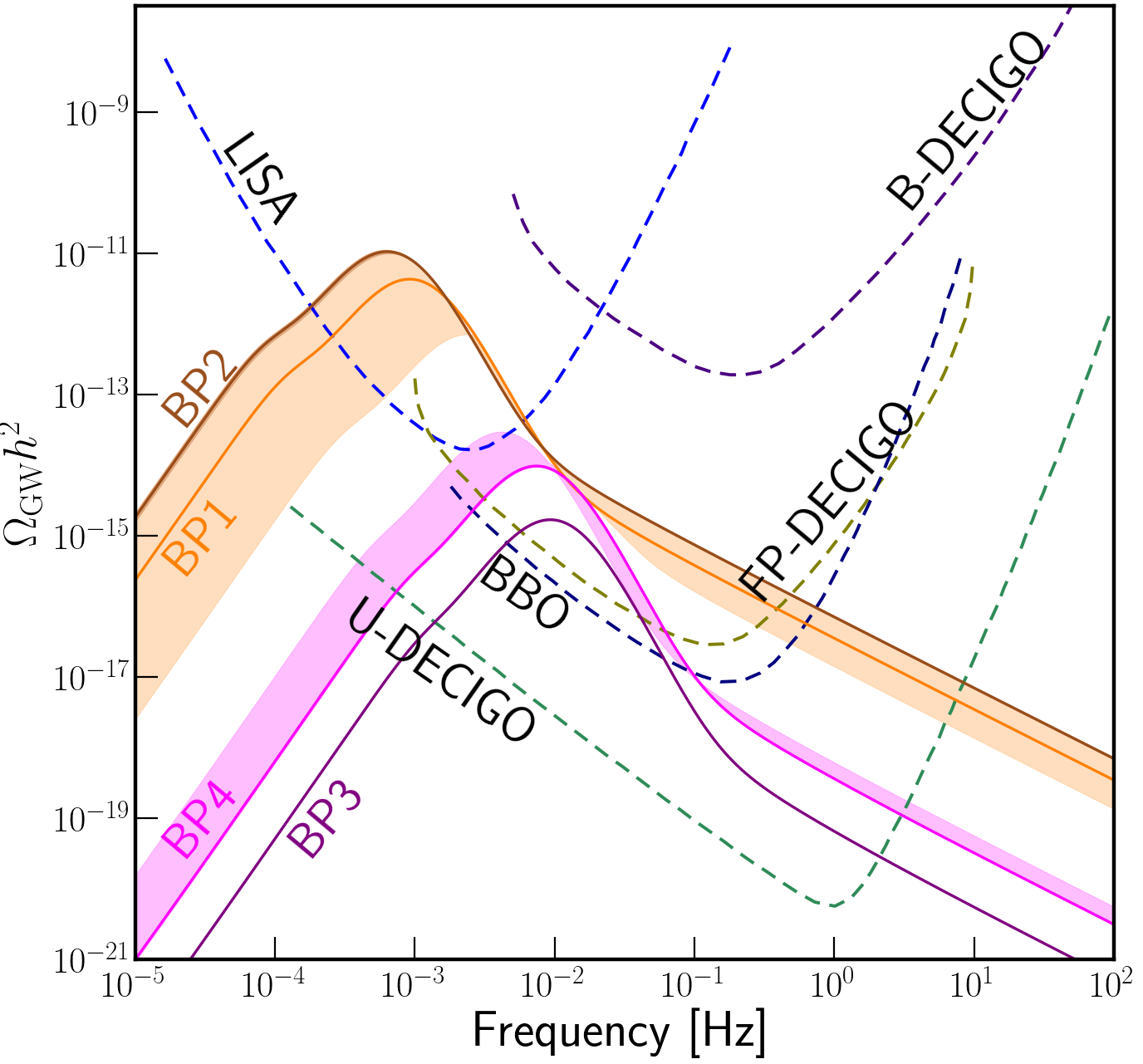} 
\caption{The maximum amplitude of the gravitational wave spectrum as a function of peak frequency (right) and stochastic gravitational wave spectrum as a function of frequency for the benchmark points (left). { The uncertainty bands for the four benchmark points are determined by varying the renormalization scale $\mu$ within the range $(0.7 m_{\psi}, 3m_{\psi})$.} We also show the experimental sensitivities for LISA~\cite{Caprini:2015zlo}, BBO~\cite{Corbin:2005ny} and three stages of  DEGICO~\cite{Kudoh:2005as} experiment.}
\label{fig:GW}
\end{figure}

The main results of our numerical analysis for the gravitational wave signals are summarized in the right panel of Fig.~\ref{fig:GW}, where we also compare the results with the future experimental sensitivities. There is a significant portion of the parameter space that generates gravitational wave signals, a sizeable portion of which will be detected by future space-based interferometers. Furthermore, we present stochastic gravitational wave spectra as a function of frequency in the left panel of~\autoref{fig:GW} for the four benchmark parameter points present in~Table \ref{tab:1}, contextualizing them with the anticipated sensitivities of several proposed space-based GW detectors. { For the four benchmark points, we also display the uncertainty bands associated with the choice of the renormalization scale $\mu^2=m_\psi^2$. These bands are obtained by varying $\mu$ within the range $(0.7 m_\psi, 3 m_\psi)$. The selection of the renormalization scale has minimal impact on our results, leading to only subpercentage differences in $\alpha$ and $T_n$. Notably, variation is primarily observed in the $\beta/H_n$ parameter. However, for BP2 and BP3, this variation is negligible. In the case of BP1 (BP4), the gravitational wave spectrum remains observable in LISA (BBO and FP-DECIGO) across all choices of the renormalization scale.} { Sound waves predominantly contribute to the gravitational wave signal. } The slight wiggle that pops up on the left side from the maximum of GW spectra is because of a small shift in the peak of GW contributions from sound waves and bubble collision. { The amplitude of the peak resulting from bubble collisions is approximately two orders of magnitude smaller than that of the sound wave contribution, and the peak frequency for bubble collisions is lower compared to the sound wave contribution.} For our study, we take $\tau=3$ years, considering that the LISA mission will last $4$ years with a duty cycle of $75\%$~\cite{Caprini:2019egz}. LISA could successfully detect BP1 and BP2 with SNR of order $(10^4-10^5)$. BP3 and BP4 could be probed in both BBO and FP-DECIGO with SNR $O(10^4)$.

 \begin{figure}
\centering
\includegraphics[width=0.68\hsize]{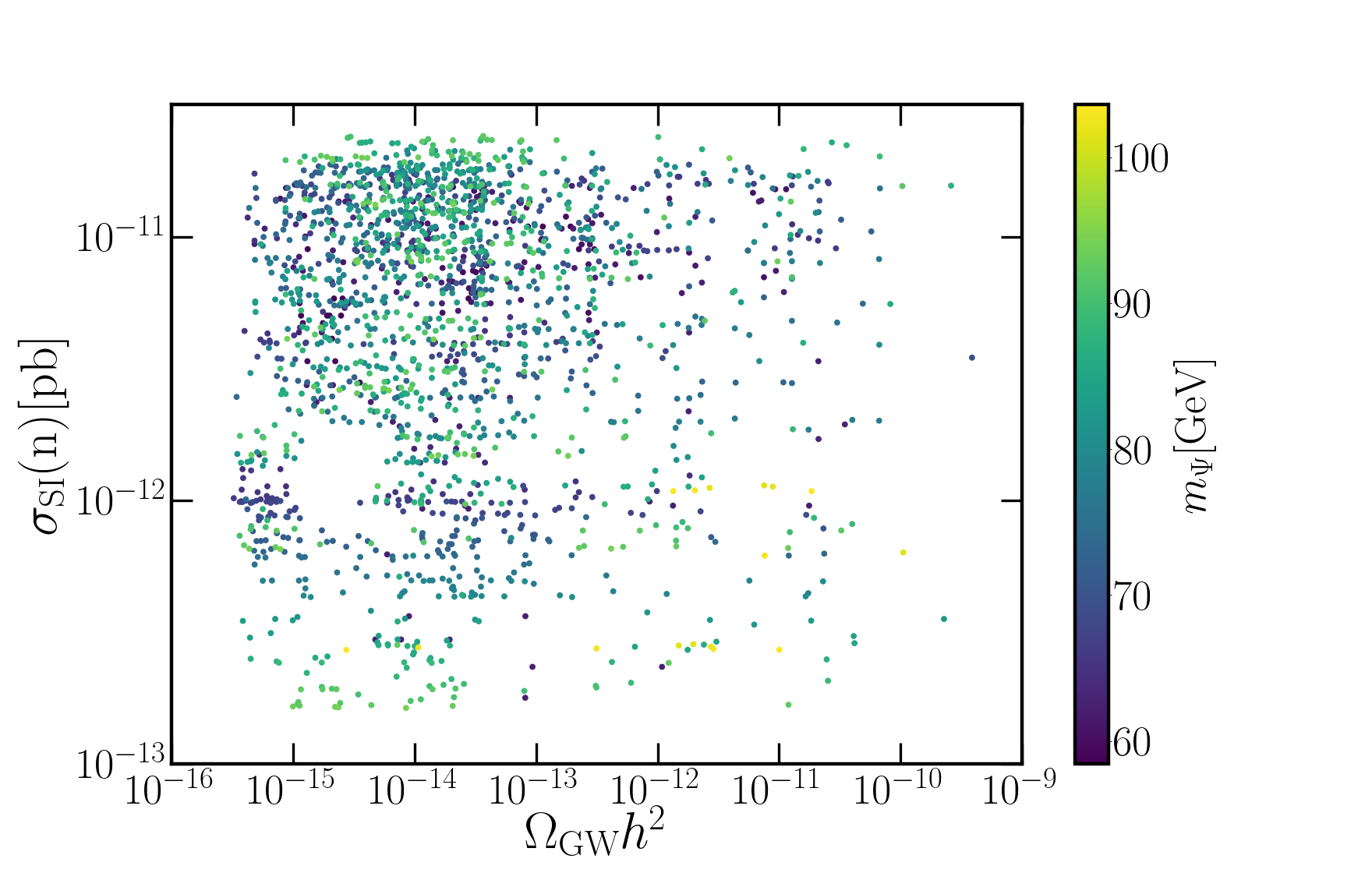} 
\caption{The maximum amplitude of the gravitational wave spectrum versus WIMP-nucleon scattering cross-section $\sigma_{\mathrm{SI}}(\mathrm{n})[\mathrm{pb}]$ color coded with dark matter mass $m_{\Psi}$. The bulk of parameter points showing detectable gravitational wave signal has a large WIMP-nucleon scattering cross-section $\sigma_\mathrm{SI}$ and could be probed in dark matter experiments.}
\label{fig:GW2}
\end{figure}

 In~\autoref{fig:GW2}, we scrutinize the potential for these parameter points to be probed in the DM direct detection experiments by showing peak amplitude of $\Omega_{\mathrm{GW}}h^2(f)$ and WIMP-nucleon scattering cross-section $\sigma_{\mathrm{SI}}(\mathrm{n})$ color coded with the DM mass. The bulk of points that give detectable GW signals in next-generation detectors have large WIMP-nucleon scattering cross-section $\sigma_\text{SI}$ and can be probed in ongoing and forthcoming DM direct detection experiments. There are parameter points that GW detectors can probe, but they are far beyond the reach of current dark-matter experiments. { Parameter points with detectable GW signal and produce correct dark matter relic abundance have DM mass in the range $58\,\mathrm{GeV}$ to $110\,\mathrm{GeV}$ and the mass of singlet scalar $m_{h^\prime}$ is close to the DM mass. Parameter points exhibiting large SNR in LISA prefer gauge coupling $g_D\in(0.5,0.7)$ and quartic coupling $\lambda_1 \lessapprox 0.2$.}


\begin{table}
\begin{tabular}{
|p{2.6cm}||p{2.5cm}|p{2.5cm}|p{2.5cm}|p{2.5cm}|  }
\hline
\multicolumn{5}{|c|}{Benchmark Points} \\
\hline
& BP1  &BP2 & BP3 & BP4 \\
\hline
\hline
\centering $m_{DM}\,\mathrm{[GeV]}$& $102.586$& $83.1724$& $76.681$& $90.5793$\\
\hline
\centering $y_{DM}$& $0.60975$& $0.539252$& $0.527997$& $0.569024$ \\
\hline
\centering $\theta_{Hh}$& $0.002$& $0.005$& $0.004$& $0.01$ \\
\hline
\centering $m_{V}\mathrm{[GeV]}$& $215.431$& $182.979$& $161.03$& $190.217$ \\
\hline
\centering $m_{H}\mathrm{[GeV]}$& $215.431$& $182.979$& $161.03$& $190.217$ \\
\hline
\centering $m_{h^\prime}\mathrm{[GeV]}$& $100.$& $80.$& $75.$& $88.$ \\   
\hline
\centering $T_{c} \mathrm{[GeV]}$& $59.0772$& $51.0072$& $51.9848$& $56.3108$\\
\hline
\centering $\xi_{c}\equiv \frac {s_c}{T_c} $& $2.757$& $2.91$& $2.65318$& $2.69087$\\
\hline
\centering $T_{n} \mathrm{[GeV]}$& $33.7387$& $27.3573$& $41.1473$& $40.320$\\
\hline
\centering $T_{p} \mathrm{[GeV]}$& $33.4634$& $27.4324$& $40.7281$& $39.7994$\\
\hline
\centering $\alpha$& $0.255441$& $0.331336$& $ 0.0610362$& $0.0989808$\\
\hline
\centering $\beta/{H_\star}$& $117.08$& $101.455$& $847.145$& $718.137$\\
\hline
\centering $\Omega h^2$& $0.12$ & $0.12$ & $0.12$ & $0.12$  \\
\hline
\centering $\sigma_\text{SI}(p)[\mathrm{pb}]$& $2.688\times10^{-13} $ & $8.59 \times 10^{-12}$ & $8.002 \times 10^{-12}$ & $1.91 \times 10^{-11}$\\
\hline
\centering $\sigma_\text{SI}(n)[\mathrm{pb}]$& $2.741 \times 10^{-13}$ &  $8.761 \times 10^{-12}$ & $8.162 \times 10^{-12}$ & $1.949 \times 10^{-11}$ \\
\hline 
\end{tabular}  
\caption{Four benchmark points for the model which has observable graviational wave signals, as presented in~Fig. \ref{fig:GW}.}
\label{tab:1}
\end{table}

\section{Summary and Conclusions}
\label{sec:conclusion}
The gravitational detectors provide an exciting new window to explore new physics after the groundbreaking observation by the LIGO and VIRGO collaborations. A fascinating source for the stochastic gravitational-wave signal is first-order phase transition due to spontaneous symmetry breaking in BSM physics. 
In this paper, we have explored such GW signals in a chiral fermion dark matter model. Such models are attractive as the dark matter mass is protected by a gauge symmetry.  Spontaneous breaking of this dark sector gauge symmetry generates the DM mass.  The universe would have gone through a phase transition associated with this dark sector symmetry breaking, which could generate observable gravitational wave signals in future experiments.

We have focused on a simple model of chiral fermion dark matter proposed in Ref.~\cite{Abe:2019zhx}.  It is based on 
a dark sector gauge symmetry $\mathrm{SU}(2)_\mathrm{D}$. A dark isospin-3/2 fermion serves as the DM candidate, which acquires its mass via the Higgs mechanism. An isospin-3 scalar field is responsible for the symmetry breaking.  We have carried out a one-loop finite-temperature effective potential analysis of this model and analyzed its phase structure as a function of the temperature.   Performing a comprehensive numerical analysis, we identified the region of parameter space that allows for first-order phase transition and computed the associated stochastic gravitational-wave signal.{\footnote{For a comprehensive analysis of the uncertainties associated with the evaluation of the GW spectrum, we refer to~\cite{Athron:2023rfq}.}} The breaking of $\mathrm{SU}(2)_L\times \mathrm{U}(1)_Y$ symmetry proceeds as in the SM. The breaking of $\mathrm{SU}(2)_\mathrm{D}$ symmetry is intriguing, as it is mostly first-order phase transition and could exhibit strong GW signals. Examining the gravitational wave peaks with the expected experimental sensitivities, we have shown that a good part of the parameter space of the model would give signals detectable by planned space-based interferometers, including LISA, BBO, and DECIGO, as shown in~\autoref{fig:GW}. To illustrate the shape of the signal, we explicitly show the GW signal for four selected benchmark points given in~\autoref{tab:1}. The SNRs calculated for the four benchmark points confirm that they could be detected by BBO, FP-DECIGO, and ULTIMATE-DECIGO experiments, and two of them could be probed in LISA. Most of these parameter points have large WIMP-nucleon scattering cross-section $\sigma_{\mathrm{SI}}$ and could be probed in dark matter experiments. This provides an exciting complementarity between gravitational wave experiments and dark matter direct detection experiments.

\section*{Acknowledgments}
The work of TA is supported in part by JSPS KAKENHI Grant Number 21K03549. The work of KSB and AK are supported in part by US Department of Energy Grant Number DE-SC 0016013.  Part of the computing for this project was performed at the High Performance Computing Center at Oklahoma State University, which is supported in part through the National Science Foundation grant OAC-1531128. 


\end{sloppypar}
\bibliographystyle{utphys}

\bibliography{ref}
\end{document}